\documentclass{llncs}

\usepackage{latexsym}

\def\rrep{ {\bf reducerep} }
\def\rep{ {\bf rep} }
\def\nrep{ {\bf normrep} }
\def\norm{ {\bf norm} }
\def\red{ {\bf reduce} }
\def\gnorm{ {\bf g\_norm}}
\def\gred{ {\bf g\_reduce}}
\def\gnrep{ {\bf g\_normrep} }
\def\grrep{ {\bf g\_reducerep} }
\def\grep{ {\bf g\_rep}}

\title{A Generalized Disjunctive Paraconsistent Data Model for Negative and Disjunctive Information}
\author{Haibin Wang, Yuanchen He and Rajshekhar Sunderraman}
\institute{Department of Computer Science \\ 
Georgia State University \\ Atlanta, GA 30302 \\
email: \{hwang17,yhe\}@student.gsu.edu, raj@cs.gsu.edu}
                                                                          
\begin{document}
\maketitle

\bibliographystyle{splncs}

\begin{abstract}
This paper presents a generalization of the disjunctive paraconsistent 
relational 
data  model in
which disjunctive positive and negative information can be represented
explicitly and manipulated. There are situations where the closed
world assumption to infer negative facts is not valid or undesirable
and there is a need to represent and reason with negation explicitly.
We consider explicit disjunctive negation in the context of disjunctive 
databases
as there is an interesting interplay between these two types
of information. {\em Generalized disjunctive paraconsistent relation}
is introduced as the main structure in this model. 
The relational algebra is appropriately
generalized to work on generalized disjunctive paraconsistent relations
and their correctness is established.
\end{abstract}

\section{Introduction}

Two important features of the relational data model \cite{cdd70} for
databases are its value-oriented nature and its rich set of simple,  
but powerful algebraic operators.
Moreover, a strong theoretical foundation for the model is provided  
by the classical first-order logic \cite{REIT84}.
This combination of a respectable theoretical platform, ease of
implementation and the practicality of the model resulted in its  
immediate success, and the model has enjoyed being used by many 
database management systems.

One limitation of the relational data model, however, is its lack of
applicability to nonclassical situations.
These are situations involving incomplete or even
inconsistent information.

Several types of incomplete information have been extensively 
studied in the past
such as {\em null} values \cite{cdd86,IL84}, 
{\em partial} values \cite{GRAN80}, {\em fuzzy} and 
{\em uncertain} values \cite{gln86,rjm88}, and 
{\em disjunctive} information \cite{LS89a,snd93}.

In this paper, we present a generalization of the disjunctive paraconsistent
data model\cite{SUN97}. 
Our model is capable of representing and
manipulating disjunctive positive facts as well as explicit disjunctive 
negative facts.
We introduce {\em generalized disjunctive paraconsistent relations}, which are 
the fundamental structures underlying our model.
These structures are generalizations of {\em disjunctive paraconsistent relations}
which are capable of representing disjunctive positive
and explicit negative definite facts. A generalized disjunctive paraconsistent 
relation
essentially consists of two kinds of information:
positive tuple sets representing exclusive disjunctive positive facts 
(one of which
belongs to the relation) and negative tuple sets representing exclusive disjunctive negated facts
(one of which does not belong to the relation).
Generalized disjunctive paraconsistent relations are strictly more general than
disjunctive paraconsistent relations in that for any disjunctive paraconsistent relation,
there is a generalized disjunctive paraconsistent relation with the same
information content, but not {\em vice versa}.
We define algebraic operators over generalized disjunctive
paraconsistent relations that extend the standard operations
over disjunctive paraconsistent relations.

\section{Motivation}

Explicit negation occurs in everyday world where
certain values cannot be given to some parameters.
In current database systems, negation is implicitly
assumed (using closed world assumption \cite{rei:cwdb}) 
when a particular query has a null answer
from the database. But this poses a problem.
Consider the following relational database 

\begin{center}
\begin{tabular}{|l|l|}
\multicolumn{2}{c}{suppliers} \\ \hline
SNUM & SNAME \\ \hline \hline
s1 & Haibin \\ \hline
s2 & Yuanchun \\ \hline
s3 & Raj \\ \hline
\end{tabular}
\hspace{.25in}
\begin{tabular}{|l|l|}
\multicolumn{2}{c}{parts} \\ \hline
PNUM & PNAME \\ \hline \hline
p1 & nut \\ \hline
p2 & cam \\ \hline
p3 & bolt \\ \hline
p4 & wheel \\ \hline
\end{tabular}
\hspace{.25in}
\begin{tabular}{|l|l|}
\multicolumn{2}{c}{supply} \\ \hline
SNUM & PNUM \\ \hline \hline
s1 & p1 \\ \hline
s1 & p3 \\ \hline
s2 & p2 \\ \hline
s3 & p4 \\ \hline
\end{tabular}
\end{center}

Consider the query ``find all suppliers who do not supply part p1 or part p3''. 
Suppose there is a known list
of suppliers, then the answer for the query would be \{ s2, s3 \}.
This may be a definite answer from the database (augmented with the CWA),
but the answer has some indefiniteness because the database may be 
{\em incomplete}. Explicit presence of incomplete information in
the form of {\em null values} complicates the problem further.
Suppose the tuple (s3,null) is part of the supply relation.
Then, we are uncertain whether to include s3 as part of the answer or not.
Finally, a similar problem occurs when one allows disjunctive
information (such as (s3,p1) or (s3,p2)) as part of the database.
 
Definite negation can occur without explicit negation in current
database systems. The use of functional dependencies provide this
facility. Consider the functional dependency that each person can 
have only one social security number. Hence if we know the SSN for 
a particular individual then we can explicitly assume the negation 
of all other possible numbers as the person's social security number.
 
Sometimes it is important to explicitly include in the database certain
negative information. Consider a medical database containing
patient information. When a doctor needs to check whether a patient has
diabetes, (s)he would be more comfortable with a negative answer
generated by the system using definite information (of explicit negative data)
than with a negative answer found using the closed world assumption.
 
We are not considering the use of negative information as an
integrity constraint. Rather we are utilizing negative information
in query processing to provide definite or disjunctive negation when needed. 
When a positive data is included in the database, which negates a previous
explicit negative information, the new data may be allowed to be entered
if the user wishes to enforce it.

In this paper, we consider explicit negation in the 
context of disjunctive databases. We extend the representation
provided in \cite{LS88a}  by introducing explicit disjunctive negative facts.
There is an interesting interplay between these two kinds of information.
Negative facts tend to reduce the amount of incompleteness
present in the disjunctive facts as seen by the equivalence
$((P \vee Q \vee R) \wedge \neg P) \equiv (Q \vee R) \wedge \neg P$.
After introducing generalized disjunctive paraconsistent relations, we present
operators to remove redundancies and inconsistencies.
We also extend the standard relational algebra to operate on
generalized disjunctive paraconsistent relations. The information content of 
generalized disjunctive paraconsistent relations is characterized in terms of 
disjucntive paraconsistent relations which we briefly present in the next 
section.

\section{Disjunctive Paraconsistent Relations}

In this section, we present a brief overview of disjunctive paraconsistent
relations and the algebraic operations on them. For a more detailed 
description, refer to \cite{SUN97}.

Let a {\em relation scheme} (or just {\em scheme})
$\Sigma$ be a finite set of {\em attribute names},
where for any attribute name
$A \in \Sigma$, $\mbox{\em dom}(A)$ is a non-empty
{\em domain} of values for $A$.
A {\em tuple} on $\Sigma$ is any map $t:\Sigma \rightarrow \cup_{A  
\in \Sigma}
\mbox{\em dom}(A)$, such that $t(A) \in \mbox{\em dom}(A)$, for each  
$A \in
\Sigma$.
Let $\tau(\Sigma)$ denote the set of all tuples on $\Sigma$.

\begin{definition}
\emph{A} paraconsistent relation \emph{on scheme} $\Sigma$ \emph{is a pair} 
$R = \langle R^+, R^- \rangle$, \emph{where} $R^+$ \emph{and} $R^-$ 
\emph{are any subsets of} $\tau(\Sigma)$. 
\emph{We let} ${\cal P}(\Sigma)$ 
\emph{be the set of all paraconsistent relations on}
$\Sigma$. \hfill{\space} $\Box$
\end{definition}

\begin{definition}
\emph{A} disjunctive paraconsistent relation, $R$, 
\emph{over the scheme}
$\Sigma$ \emph{consists of two components} $<R^{+}, R^{-}>$
\emph{where} $R^{+} \subseteq 2^{\tau(\Sigma)}$ \emph{and}
$ R^{-} \subseteq \tau(\Sigma)$.
$R^{+}$, \emph{the} positive \emph{component, is a set of tuple sets}.
\emph{Each tuple set in this component represents a disjunctive} 
\emph{positive fact. In the case where the} 
\emph{tuple set is a singleton, we have a definite positive fact.} 
$R^{-}$, \emph{the} negative \emph{component consists} 
\emph{of tuples that we refer to as definite negative tuples.} 
\emph{Let} ${\cal D}(\Sigma)$ \emph{represent all disjunctive paraconsistent}
\emph{relations over the scheme} $\Sigma$.
\hfill{\space} $\Box$
\end{definition}

\begin{definition}
\emph{Let} $R$ \emph{be a disjunctive paraconsistent relation over} $\Sigma$. 
\emph{Then}, \\
$\norm(R)^{+} = \{ w | w \in R^{+} \wedge w \not \subseteq R^{-}\}$\\
$\norm(R)^{-} = R^{-} - \{t | t \in R^{-} \wedge (\exists w)(
			      w \in R^{+} \wedge t \in w \wedge 
			      w \subseteq R^{-}) 
			\}$
\hfill{\space} $\Box$
\end{definition}
A disjunctive paraconsistent relation is called {\em normalized}
if it does not contain any inconsistencies.
We let ${\cal N}(\Sigma)$ denote the set of all normalized disjunctive
paraconsistent relations over scheme $\Sigma$.

\begin{definition}
\emph{Let} $R$ \emph{be a normalized disjunctive paraconsistent}
\emph{relation. Then},  $\red(R)$ \emph{is defined as follows}:\\
$\red(R)^{+} = \{ w' \mid (\exists w) (w \in R^{+} \wedge 
                           w' = w - R^{-} \wedge$ \\
\hspace*{1.2in} $\neg (\exists w_1) (w_1 \in R^{+} \wedge 
		       (w_1 - R^{-}) \subset w') ) \}$ \\
$\red(R)^{-} = R^{-}$  
\hfill{\space} $\Box$
\end{definition}

\begin{definition}
\emph{Let} $U \subseteq {\cal P}(\Sigma)$. \emph{Then}, \\
$\nrep_{\Sigma}(U) = 
U - \{ R | R \in U \wedge R^{+} \cap R^{-} \neq \emptyset \}$
\hfill{\space} $\Box$
\end{definition}

The $\nrep$ operator removes all inconsistent paraconsistent relations
from its input.
 
\begin{definition}
\emph{Let} $U \subseteq {\cal P}(\Sigma)$. \emph{Then}, 
$$\rrep_{\Sigma}(U) = \{ R | R \in U \wedge \neg (\exists S)(S \in U \wedge 
                   R \neq S \wedge S^{+} \subseteq R^{+} \wedge 
                   S^{-} \subseteq R^{-}) \}$$
\hfill{\space} $\Box$
\end{definition}

The $\rrep$ operator keeps only the ``minimal'' paraconsistent
relations and eliminates any paraconsistent relation that is ``subsumed''
by others.

\begin{definition}
\emph{The information content of disjunctive paraconsistent relations}
\emph{is defined by the mapping} $\rep_{\Sigma}: {\cal N}(\Sigma) \rightarrow
{\cal P}(\Sigma)$. \emph{Let}
$R$ \emph{be a normalized disjunctive paraconsistent relation}
\emph{on scheme} $\Sigma$ \emph{with} $R^{+} = \{ w_{1}, \ldots, w_{k} \}$.
\emph{Let}
$U = \{ <\{t_{1}, \ldots, t_{k}\},R^{-}> | (\forall i)(1 \leq i \leq k 
\rightarrow t_{i} \in w_{i}) \}$. 
\emph{Then}, \\
$\rep_{\Sigma}(R) = \rrep_{\Sigma}(\nrep_{\Sigma}(U))$
\hfill{\space} $\Box$
\end{definition}

\begin{definition}
\emph{Let} $R$ \emph{and} $S$ \emph{be two normalized disjunctive}
\emph{paraconsistent relations on scheme}
$\Sigma$ \emph{with} $\red(R)^+ = \{v_1, \ldots, v_n\}$ \emph{and}
$\red(S)^+ = \{w_1, \ldots, w_m\}$. \emph{Then}, $R \widehat{\cup} S$
\emph{is a disjunctive paraconsistent relation over scheme} $\Sigma$ 
\emph{given by}
$R \widehat{\cup} S = \red(T)$, \emph{where}
$T^{+} = \red(R)^{+} \cup \red(S)^{+}$ and
$T^{-} = \red(R)^{-} \cap \red(S)^{-}$. 

\noindent
\emph{and} $R \widehat{\cap} S$
\emph{is a disjunctive paraconsistent relation over scheme} $\Sigma$ 
\emph{given by}
$R \widehat{\cap} S = \red(T)$, \emph{where} $T$ 
\emph{is defined as follows}. 
\emph{Let} $E = \{\{t_1, \ldots, t_n\}|(\forall i)(1 \leq i \leq n \rightarrow t_i \in v_i)\}$ \emph{and} $F = \{\{t_1, \ldots, t_m\}|(\forall i)(1 \leq i \leq m \rightarrow t_i \in w_i)\}$. \emph{Let the elements of} $E$ \emph{be}
$E_1, \ldots, E_e$ \emph{and those of} $F$ \emph{be} $F_1, \ldots, F_f$ 
\emph{and let} $A_{ij} = E_i \cap F_j$ \emph{for} $1 \leq i \leq e$ \emph{and}
$1 \leq j \leq f$. \emph{Let} $A_1, \ldots, A_g$ \emph{be the distinct
} $A_{ij}$s. \emph{Then}, \\
$T^{+} = \{w|(\exists t_1) \cdots (\exists t_g)(t_1 \in A_1 \wedge \cdots \wedge t_g \in A_g \wedge w = \{t_1, \ldots, t_g\})\}$ \\ 
$T^{-} = R^{-} \cup S^{-}$. 
\hfill{\space} $\Box$
\end{definition}

\begin{definition}
\emph{Let} $R$ 
\emph{be a normalized disjunctive paraconsistent relation on scheme} 
$\Sigma$,
\emph{and let} $F$ \emph{be any logic formula involving attribute names in}  
$\Sigma$,
\emph{constant symbols}
\emph{(denoting values in the attribute domains), equality symbol} $=$,
\emph{negation symbol} $\neg$, \emph{and connectives} $\vee$ \emph{and}
$\wedge$.
\emph{Then, the} selection \emph{of} $R$ \emph{by} $F$,
\emph{denoted} $\widehat{\sigma}_F(R)$, 
\emph{is a} 
\emph{disjunctive paraconsistent relation on scheme}  $\Sigma$, \emph{given by}
$\widehat{\sigma}_{F}(R) = \red(T)$, \emph{where}
$T^{+} = \{ w | w \in \red(R)^{+} \wedge (\forall t \in w) F(t) \}$ \emph{and}
$T^{-} = \red(R)^{-} \cup \sigma_{\neg F}(\tau(\Sigma))$, 
\emph{where} $\sigma_F$ \emph{is the usual selection of tuples}.
\hfill{\space} $\Box$
\end{definition}

If $\Sigma$ and $\Delta$ are relation schemes such that $\Sigma \subseteq \Delta$, then for any tuple $t \in \tau(\Sigma)$, we let $t^\Delta$ denote the set $\{t' \in \tau(\Delta) ~|~ t'(A) = t(A) \mbox{, for all $A \in \Sigma$}\}$ of all extensions of $t$. We extend this notion for any $T \subseteq \tau(\Sigma)$ by defining $T^\Delta = \cup_{t \in T} ~ t^\Delta$.

\begin{definition}
\emph{Let} $R$ \emph{be a normalized disjunctive paraconsistent relation} 
\emph{on scheme} $\Sigma$, \emph{and} $\Delta \subseteq \Sigma$.
\emph{Then, the} projection \emph{of} $R$ \emph{onto} $\Delta$,
\emph{denoted} $\widehat{\pi}_\Delta(R)$, \emph{is a disjunctive}
\emph{paraconsistent relation on  scheme} $\Delta$, \emph{given by}
$\widehat{\pi}_{\Delta}(R) = \red(T)$, \emph{where}
$T^{+} = \{ \pi_{\Delta}(w) | w \in \red(R)^{+} \}$ \emph{and}
$T^{-} = \{ t \in \tau(\Delta) |  t^{\Sigma \cup \Delta}
\subseteq$ \\
$(\red(R)^-)^{\Sigma \cup \Delta} \}$,
\emph{where} $\pi_\Delta$ 
\emph{is the usual projection over $\Delta$ of tuples.}
\hfill{\space} $\Box$
\end{definition}

\begin{definition}
\emph{Let} $R$ \emph{and} $S$ 
\emph{be normalized disjunctive paraconsistent relations on schemes} 
$\Sigma$ \emph{and}  $\Delta$, \emph{respectively with} 
$\red(R)^{+} = \{ v_{1}, \ldots, v_{n} \}$ \emph{and}
$\red(S)^{+} = \{ w_{1}, \ldots, w_{m} \}$.
\emph{Then, the} natural join \emph{of} $R$ \emph{and} $S$,
\emph{denoted} $R~\widehat{\Join}~S$, 
\emph{is a disjunctive paraconsistent relation on scheme}
$\Sigma \cup \Delta$,
\emph{given by} $R~\widehat{\Join}~S = \red(T)$, 
\emph{where} $T$ \emph{is defined as follows. Let}
$E = \{ \{ t_{1}, \ldots, t_{n} \} | (\forall i)(1 \leq i \leq n \rightarrow
t_{i} \in v_{i}) \}$ \emph{and} 
$F = \{ \{ t_{1}, \ldots, t_{m} \} | (\forall i)(1 \leq i \leq m \rightarrow
t_{i} \in w_{i}) \}$.
\emph{Let the elements of} $E$ \emph{be} $E_{1}, \ldots, E_{e}$
\emph{and those of} $F$ \emph{be} $F_{1}, \ldots, F_{f}$ \emph{and let} 
$A_{ij} = E_{i} \Join F_{j} $
\emph{for} $1 \leq i \leq e$ \emph{and} $1 \leq j \leq f$. \emph{Let}
$A_{1}, \ldots, A_{g}$ \emph{be the distinct} $A_{ij}$\emph{s}. 
\emph{Then},

\begin{tabular}{l}
$T^{+} = \{ w | (\exists t_{1}) \cdots (\exists t_{g}) (
t_{1} \in A_{i} \wedge \cdots \wedge t_{g} \in A_{g} \wedge
w = \{ t_{1}, \ldots, t_{g} \} ) \}$ \\
$T^- = (\red(R)^-)^{\Sigma \cup \Delta} \cup (\red(S)^-)^{\Sigma \cup \Delta}.$
\end{tabular}

\hfill{\space} $\Box$
\end{definition}

\section{Generalized Disjunctive Paraconsistent Relations}
In this section, we present the main structure underling our model, the
{\em generalized disjunctive paraconsistent relations}. We identify several
types of redundancies and inconsistencies that may appear and provide operators
to remove them. Finally, we present the information content of generalized
paraconsistent relations.

\begin{definition}
\emph{A} generalized disjunctive paraconsistent relation, $R$, \emph{over the scheme}
$\Sigma$ \emph{consists of two components} $\langle R^+,R^- \rangle$ \emph{where} 
$R^+ \subseteq 2^{\tau(\Sigma)}$ \emph{and} $R^- \subset 2^{\tau(\Sigma)}$. $R^+$,
\emph{the} positive \emph{component, is a set of tuple sets. Each tuple set in this
component represens a disjunctive positive fact. In the case where the tuple
set is a singleton, we have a definite positive fact}. $R^-$, \emph{the} 
negative
\emph{component consists of a set of tuple sets. Each tuple set in this component
represents a disjunctive negative fact. In the case where the tuple set is a
singleton, we have a definite negated fact. Let} ${\cal {GD}}(\Sigma)$ \emph{represent
all generalized disjunctive paraconsistent relatios over the scheme} $\Sigma$.
\hfill{\space} $\Box$
\end{definition}

\begin{example}
Consider the following generalized disjunctive paraconsistent relation:\\
$supply^{+} = \{\{<s1,p1> \},\{<s2,p1>, <s2,p2> \},\{<s3,p3>, <s3,p4> \} \}$ \\
$supply^{-} = \{\{<s1,p2> \}, \{<s1,p3>\}, \{<s2,p3>, <s2,p4> \}\}$.
The {\em positive component} corresponds to the statement $s1$ supplies 
$p1$, $s2$ supplies $p1$ or $p2$, and $s3$ supplies $p3$ or $p4$ and the 
{\em negative component} corresponds to $s1$ does not supply $p2$ and $s1$ 
does not supply $p3$ and $s2$ does not supply $p3$ or $s2$ does not 
supply $p4$. It should be noted that the status of tuples that do not appear 
anywhere in the generalized disjunctive paraconsistent relation, such as 
$(s3,p2)$, is unknown.
\hfill{\space} $\Box$
\end{example}

Inconsistences can be present in a genearlaized disjunctive paraconsistent
relation in two situations. On the one hand, if all the tuples of a tuple set 
of the posistive component are also present in the union of the singleton tuple
set of the negative component. In such a case, the tuple set states that at 
least one of the tuples in the tuple set must be in the relation whereas the 
negative component states that all the tuples in the tuple set must not be in
the relation. We deal with this inconsistency by removing both the positive
tuple set and all its corresponding singleton tuple sets from the negative
component. On the other hand, if all the tuples of a tuple set of the negative
component are also present in the union of the singleton tuple set of the 
positive component. In such a case, the tuple set states that at least one of
the tuples in the tuple set must not be in the relation whereas the positive
component states that all the tuples in the tuple set must be in the relation.
We deal with this inconsistency by removing both the negative tuple set and
all its corresponding singleton tuple sets from the positive component. This
is done by the $\gnorm$ operator defined as follows:

\begin{definition} 
\emph{Let} $R$ \emph{be a generalized disjunctive paraconsistent relation over} $\Sigma$. 
$R^+ = \{w_1, w_2, \cdots, w_n \}$ \emph{and} $R^- = \{u_1, u_2, \cdots, u_m \}$.
\emph{Then}, \\
$\gnorm(R)^{+} = R^+ - \\ \hspace*{0.5in} 
\{w | w \in R^+ \wedge w \subseteq \cup u_i \wedge 1 \leq i \leq m \rightarrow u_i \in R^- \wedge |u_i| = 1\} - \\ \hspace*{0.5in}
\{w_i | 1 \leq i \leq n \rightarrow w_i \in R^+ \wedge |w_i| = 1 \wedge (\exists u)(u \in R^- \wedge u \subseteq \cup w_i \wedge w_i \subseteq u)\}$ \\
$\gnorm(R)^{-} = R^- - \\ \hspace*{0.5in}
\{u | u \in R^- \wedge u \subseteq \cup w_i \wedge 1 \leq i \leq n \rightarrow  
w_i \in R^+ \wedge |w_i| = 1\} - \\ \hspace*{0.5in}
\{u_i | 1 \leq i \leq m \rightarrow u_i \in R^- \wedge |u_i| = 1 \wedge (\exists w)(w \in R^+ 
\wedge w \subseteq \cup u_i \wedge u_i \subseteq w)\}$  

\hfill{\space} $\Box$
\end{definition}

A generalized disjunctive paraconsistent relation is called {\em normalized} if
it does not contain any inconsistencies. We let $\cal GN(\Sigma)$ denote the set
of all normalized generalized disjunctive paraconsistent relations over scheme
$\Sigma$.
We now identify the following four types of redundancies in a normalized 
generalized disjunctive paraconsistent relation $R$: \\

\vspace{-.1in}

\begin{enumerate}
\item \underline{$w_1 \in R^+$, $w_2 \in R^+$, and $w_1 \subset w_2$.}
In this case, $w_1$ subsumes $w_2$. To eliminate this redundancy, we delete
$w_2$ from $R^+$.

\item \underline{$u_1 \in R^-$, $u_2 \in R^-$, and $u_1 \subset u_2$.}
In this case, $u_1$ subsumes $u_2$. To eliminate this redundancy, we delete
$u_2$ from $R^-$.

\item \underline{$1 \leq i \leq n$, $w_i \in R^+$, $|w_i| = 1$, $u \in R^-$, and $\cup w_i \subset u$.} This redundancy is eliminated by deleting the tuple set
$u$ from $R^-$ and adding the tuple set $u - \cup w_i$ to $R^-$. Since we are
dealing with normalized generalized disjunctive paraconsistent relations,
$u - \cup w_i$ cannot be empty.

\item \underline{$1 \leq i \leq m$, $u_i \in R^-$, $|u_i| = 1$, $w \in R^+$, and $\cup u_i \subset w$.} This redundancy is eliminated by deleting the tuple set
$w$ from $R^+$ and adding the tuple set $w - \cup u_i$ to $R^+$. Since we are
dealing with normalized generalized disjunctive paraconsistent relations,
$w - \cup u_i$ cannot be empty. 
\end{enumerate}
We now introduce an operator called $\gred$ to take care of redundancies.

\begin{definition}
\emph{Let} $R$ \emph{be a normalized generalized disjunctive paraconsistent
relation. Then,} \\
$\gred(R)^+ = \{w' | (\exists w) (w \in R^+ \wedge w' = w - U \wedge$ \\
\hspace*{1.2in} $\neg (\exists w_1) (w_1 \in R^+ \wedge (w_1 - U) \subset w')) \}$ \\
$\gred(R)^- = \{u' | (\exists u) (u \in R^- \wedge u' = u - W \wedge$ \\
\hspace*{1.2in} $\neg (\exists u_1) (u_1 \in R^- \wedge (u_1 -W) \subset u')) \}$ \\
\emph{where}, $U = \{u_i | u_i \in R^- \wedge |u_i| = 1\}$ \emph{and} $W = \{w_i | w_i \in R^+ \wedge |w_i| = 1\}$.
\hfill{\space} $\Box$ 
\end{definition}

\begin{example}
Consider the following generalized disjunctive paraconsistent relation:
$R^+ = \{ \{<a>\}, \{<b>,<c>\}, \{<c>,<d>\}, \{<a>,<e>\},\{<f>,<g>\} \}$ \\
and $R^- = \{ \{<b>\}, \{<c>,<e>\}, \{<i>\}, \{<d>,<e>,<f>\} \}$.
The disjunctive tuple $\{<a>,<e>\}$ is subsumed by $\{<a>\}$ and hence removed.
In the disjunctive tuple set $\{<b>,<c>\}$, $<b>$ is redundant due to the 
presence of the negative singleton tuple set $\{<b>\}$ resulting in the positive tuple $\{<c>\}$ which in turn subsumes $\{<c>,<d>\}$ and makes $\{<c>,<e>\}$
redundant and resulting in $\{<e>\}$ which subsumes the $\{<d>,<e>,<f>\}$.
The reduced generalized disjunctive paraconsistent relation is:
$\gred(R)^+ = \{ \{<a>\}, \{<c>\}, \{<f>,<g>\} \}$ and
$\gred(R)^- = \{ \{<b>\}, \{<e>\}, \{<i>\} \}$
\hfill{\space} $\Box$
\end{example}

The information content of a generalized disjunctive paraconsistent relation
can be defined to be a collection of disjunctive paraconsistent relations.
The different possible disjunctive paraconsistent relations are constructed
by selecting one of the several tuples within a tuple set for each tuple set
in the the negative component. In doing so, we may end up with non-minimal
disjunctive paraconsistent relations or even with inconsistent disjunctive
paraconsistent relations. These would have to be removed in order to obtain
the exact information content of generalized disjunctive paraconsistent
relations. The formal definitions follow:

\begin{definition}
\emph{Let} $U \subseteq {\cal D}(\Sigma)$. \emph{Then},
$\gnrep_\Sigma(U) = \{R | R \in U \wedge \neg (\exists w)(w \in R^+ \wedge w \subseteq R^-)\}$
\hfill{\space} $\Box$
\end{definition}

The $\gnrep$ operator removes all inconsistent disjunctive paraconsistent 
relations from its input.

\begin{definition}
\emph{Let} $U \subseteq {\cal D}(\Sigma)$. \emph{Then},
$\grrep_\Sigma(U) = \{R | R \in U \wedge \neg (\exists S)(S \in U \wedge R \neq S \wedge S^+ \subseteq R^+ \wedge S^- \subseteq R^-)\}$
\hfill{\space} $\Box$
\end{definition}

The $\grrep$ operator keeps only the ``minimal'' disjunctive paraconsistent relations and eliminates any disjunctive paraconsistent relation that is ``subsumed'' by others.

\begin{definition}
\emph{The information content of generalized disjunctive paraconsistent 
relations is defined by the mapping} $\grep_\Sigma ~:~ {\cal {GN}}(\Sigma) \rightarrow {\cal D}(\Sigma)$. \emph{Let} $R$ \emph{be a normalized generalized disjunctive
paraconsistent relation on scheme} $\Sigma$ with $R^- = \{u_1, \ldots, u_m\}$.
\emph{Let} $U = \{R^+, <\{t_1, \ldots, t_m\}> | (\forall i)(1 \leq i \leq m \rightarrow t_i \in u_i)\}$. \emph{Then},
$\grep_\Sigma(R) = \grrep_\Sigma(\gnrep_\Sigma(U))$
\hfill{\space} $\Box$ 
\end{definition}

Note that the information content is defined only for normalized generalized 
disjunctive paraconsistent relations.

\begin{example}
Consider the following generalized disjunctive paraconsistent relation on a
single attribute scheme $\Sigma$:
$R^+ = \{ \{<b>,<e>\},\{<c>,<d>\},\{<e>,<g>\} \} and R^- = \{ \{<b>\}, \{<c>,<e>\}, \{<c>, <d>,<g>\}\}$
The process of selecting tuples from tuple sets produces the following 
disjunctive paraconsistent relations:

$U = \{ <\{\{ \{<b>,<e>\}, \{<c>,<d>\}, \{<e>,<g>\}\}, \{<b>,<c>\} \} >, 
        <\{\{ \{<b>,<e>\}, \{<c>,<d>\}, \{<e>,<g>\}\}, \{<b>,<c>,<d>\} \} >,
        <\{\{ \{<b>,<e>\}, \{<c>,<d>\}, \{<e>,<g>\}\}, \{<b>,<c>,<g>\} \} >, 
        <\{\{ \{<b>,<e>\}, \{<c>,<d>\}, \{<e>,<g>\}\}, \{<b>,<e>,<c>\} \} >,
        <\{\{ \{<b>,<e>\}, \{<c>,<d>\}, \{<e>,<g>\}\}, \{<b>,<e>,<d>\} \} >,
        <\{\{ \{<b>,<e>\}, \{<c>,<d>\}, \{<e>,<g>\}\}, \{<b>,<e>,<d>\} \} > \}$.

Normalizing the above set of disjunctive paraconsistent relations using \\
$\gnrep$ gives us:
$U' = \{<\{\{ \{<b>,<e>\}, \{<c>,<d>\}, \{<e>,<g>\}\}, \{<b>,<c>\} \} >,
        <\{\{ \{<b>,<e>\}, \{<c>,<d>\}, \{<e>,<g>\}\}, \{<b>,<c>,<g>\} \} >\}$. 

Finally, removing the non-minimal disjunctive paraconsistent relations using 
the $\grrep$ operator, we get the information content $\grep_\Sigma(R)$ as 
follows:
$\grep_\Sigma(R) = \{<\{\{ \{<b>,<e>\}, \{<c>,<d>\}, \{<e>,<g>\}\}, \{<b>,<c>\} \}>\}$.
\hfill{\space} $\Box$        
\end{example}

The following important theorem states that information is neither lost nor 
gained by removing the redundancies in a generalized disjunctive paraconsistent
relations.

\begin{theorem}
Let $R$ be a generalized disjunctive paraconsistent relation on
scheme $\Sigma$. Then, \\
$\grep_{\Sigma}(\gred(R)) = \grep_{\Sigma}(R)$
\hfill{\space} $\Box$
\end{theorem}

\section{Generalized Relational Algebra}
In this section, we first develop the notion of {\em precise generalizations}
of algebraic operators. This is an important property that must be satisfied
by any new operator defined for generalized disjunctive paraconsistent 
relations. Then, we present several algebraic operators on generalized
disjunctive paraconsistent relations that are precise generalizations of their
counterparts on disjunctive paraconsistent relations.

\subsection*{Precise Generalization of Operations}

It is easily seen that generalized disjunctive paraconsistent relations are a 
generalization of disjunctive paraconsistent relations, in that for each 
disjunctive paraconsistent relation there is a generalized disjunctive
paraconsistent relation with the same information content, but not 
{\em vice versa}. It is thus natural to think of generalising the operations
on disjunctive paraconsistent relations, such as union, join, projection etc.,
to generalized disjunctive paraconsistent relations. However, any such 
generalization should be intuitive with respect to the {\em belief system}
model of generalized disjunctive paraconsistent relations. We now construct a
framework for operators on both kinds of relations and introduce the notion of
the precise generalization relationship among their operators.

An $n$-ary {\em operator on disjunctive paraconsistent relations with 
signature} \\ $\langle \Sigma_1, \ldots, \Sigma_{n+1} \rangle$ is a function $\Theta~:~{\cal D}(\Sigma_1) \times \cdots \times {\cal D}(\Sigma_n) \rightarrow {\cal D}(\Sigma_{n+1})$, where $\Sigma_1, \ldots, \Sigma_{n+1}$ are any schemes. Similarly, an $n$-ary
{\em operator on generalized disjunctive paraconsistent relations with signature} $\langle \Sigma_1, \ldots, \Sigma_{n+1} \rangle$ is a function: $\Psi: {\cal {GD}}(\Sigma_1) \times \cdots \times {\cal {GD}}(\Sigma_n) \rightarrow {\cal {GD}}(\Sigma_{n+1})$.

We now need to extend operators on disjunctive paraconsistent relations to 
sets of disjunctive paraconsistent relations. For any operator $\Theta~:~{\cal 
D}(\Sigma_1) \times \cdots \times {\cal D}(\Sigma_n) \rightarrow {\cal D}(\Sigma
_{n+1})$ on disjunctive paraconsistent relations, we let ${\cal S}(\Theta)~:~ 2^{{\cal {D}}(\Sigma_1)} \times \cdots \times 2^{{\cal {D}}(\Sigma_n)} \rightarrow 2^{{\cal {D}}(\Sigma_{n+1})}$ be a map on sets of disjunctive paraconsistent 
relations
defined as follows. For any sets $M_1, \ldots, M_n$ of disjunctive 
paraconsistent
relations on schemes $\Sigma_1, \ldots, \Sigma_n$, respectively, \\

${\cal S}(\Theta)(M_1, \ldots, M_n) = \{\Theta(R_1, \ldots, R_n) | R_i \in M_i, \mbox{for all } i, 1 \leq i \leq n\}$. \\

In other words, ${\cal S}(\Theta)(M_1, \ldots, M_n)$ is the set of $\Theta$-images of all tuples in the cartesian product $M_1 \times \cdots \times M_n$. We
are now ready to lead up to the notion of precise operator generalization.

\begin{definition}
\emph{An operator} $\Psi$ \emph{on generalized disjunctive paraconsistent
relations with signature} $\langle \Sigma_1, \ldots, \Sigma_{n+1} \rangle$ 
\emph{is} consistency preserving \emph{if for any normalized generalized 
disjunctive relations} $R_1, \ldots, R_n$ \emph{on schemes} $\Sigma_1, \ldots,
\Sigma_n$, \emph{respectively}, $\Psi(R_1, \ldots, R_n)$ \emph{is also 
normalized}.
\hfill{\space} $\Box$
\end{definition}

\begin{definition}
\emph{A consistency preserving operator} $\Psi$ \emph{on generalized disjunctive paraconsistent relations with signature} 
$\langle \Sigma_1, \ldots, \Sigma_{n+1} \rangle$ 
\emph{is a} precise generalization \emph{of an operator} $\Theta$ \emph{on disjunctive paraconsistent relations with the same signature, if for any normalized 
generalized disjunctive paraconsistent relations} $R_1, \ldots, R_n$ 
\emph{on schemes} $\Sigma_1, \ldots, \Sigma_n$, \emph{we have} \\ 
$\grep_{\Sigma_{n+1}}(\Psi(R_1, \ldots, R_n)) = {\cal S}(\Theta)(\grep_{\Sigma_1}(R_1), \ldots, \grep_{\Sigma_n}(R_n))$.

\hfill{\space} $\Box$
\end{definition}

We now present precise generalizations for the usual relation operators, such as union, join, projection. To reflect generalization, a line is placed over an
ordinary operator. For example, $\Join$ denotes the natural join among ordinary
relations, $\dot{\Join}$ denotes natural join on paraconsistent relations,
$\widehat{\Join}$ denotes natural join on disjunctive paraconsistent relations
and $\overline{\Join}$ denotes natural join on generalized disjunctive 
paraconsistent relations.

\begin{definition}
\emph{Let} $R$ \emph{and} $S$ \emph{be two normalized generalized disjunctive
paraconsistent relations on scheme} $\Sigma$ \emph{with} $\gred(R)^+ = \{v_1, \ldots, v_n\}$, \\
$\gred(R)^- = \{u_1, \ldots, u_k\}$ \emph{and} $\gred(S)^+ = \{w_1, \ldots, w_m\}$, \\
$\gred(S)^- = \{x_1, \ldots, x_j\}$. \emph{Then}, 
$R \overline{\cup}S$
is a generalized disjunctive paraconsistent relation over scheme $\Sigma$ 
given by $R \overline{\cup} S = \gred(T)$, \emph{where} $T$ \emph{is defined
as follows}. \emph{Let} $E = \{\{t_1, \ldots, t_k\}|(\forall i)(1 \leq i \leq k \rightarrow t_i \in u_i)\}$ \emph{and} $F = \{\{t_1, \ldots, t_j\}|(\forall i)(1 \leq i \leq j \rightarrow t_i \in x_i)\}$. \emph{Let the elements of} $E$ \emph{be} $E_1, \ldots, E_e$ \emph{and those of} $F$ \emph{be} $F_1, \ldots, F_f$
\emph{and let} $A_{ij} = E_i \cap F_j$, \emph{for} $1 \leq i \leq e$ \emph{and}
$1 \leq j \leq f$. \emph{Let} $A_1, \ldots, A_g$ \emph{be the distinct} $A_{ij}$s. \emph{Then}, \\ 
$T^+ = \gred(R)^+ \cup \gred(S)^+$ \\ 
$T^- = \{w|(\exists t_1) \cdots (\exists t_g)(t_1 \in A_1 \wedge \cdots \wedge t_g \in A_g \wedge w = \{t_1, \ldots, t_g\})\}$. \\ 
\emph{and} $R \overline{\cap} S$ is a 
generalized disjunctive paraconsistent relation over scheme $\Sigma$ given by
$R \overline{\cap} S = \gred(T)$, \emph{where} $T$ \emph{is defined
as follows}. \\
\emph{Let} $E = \{\{t_1, \ldots, t_n\}|(\forall i)(1 \leq i \leq n 
\rightarrow t_i \in v_i)\}$ \emph{and} $F = \{\{t_1, \ldots, t_m\}|(\forall i)(1
 \leq i \leq m \rightarrow t_i \in w_i)\}$. \emph{Let the elements of} $E$ \emph
{be} $E_1, \ldots, E_e$ \emph{and those of} $F$ \emph{be} $F_1, \ldots, F_f$
\emph{and let} $A_{ij} = E_i \cap F_j$, \emph{for} $1 \leq i \leq e$ \emph{and}
$1 \leq j \leq f$. \emph{Let} $A_1, \ldots, A_g$ \emph{be the distinct} $A_{ij}$
s. \emph{Then}, \\
$T^+ = \{w|(\exists t_1) \cdots (\exists t_g)(t_1 \in A_1 \wedge \cdots \wedge t
_g \in A_g \wedge w = \{t_1, \ldots, t_g\})\}$. \\
$T^- = \gred(R)^- \cup \gred(S)^-$.
\hfill{\space} $\Box$
\end{definition}

The following theorem establishes the {\em precise generalization} property 
for union and intersection:

\begin{theorem}
Let $R$ and $S$ be two normalized generalized disjunctive
paraconsistent relations on scheme $\Sigma$. Then,

\vspace{-.1in}

\begin{enumerate}
\item $\grep_{\Sigma}(R \overline{\cup} S) = \grep_{\Sigma}(R) {\cal S}(\widehat{\cup})\grep_{\Sigma}(S)$.
\item $\grep_{\Sigma}(R \overline{\cap} S) = \grep_{\Sigma}(R) {\cal S}(\widehat{\cup})\grep_{\Sigma}(S)$.
\hfill{\space} $\Box$
\end{enumerate}
\end{theorem}

\begin{definition}
\emph{Let} $R$ \emph{be normalized generalized disjunctive paraconsistent
relation on scheme} $\Sigma$. \emph{Then}, $\overline{-}R$ \emph{is a
generalized
disjunctive paraconsistent relation over scheme} $\Sigma$ \emph{given by} \\
 $(\overline{-}
R)^+ = \gred(R)^-$ \emph{and} $(\overline{-}R)^- =  \gred(R)^+$.

\hfill{\space} $\Box$
\end{definition}

\begin{definition}
\emph{Let} $R$ \emph{be a normalized generalized disjunctive paraconsistent 
relation on scheme} $\Sigma$, \emph{and let} $F$ \emph{be any logic formula 
involving attribute names in} $\Sigma$, \emph{constant symbols (denoting 
values in the attribute domains), equality symbol} $=$, \emph{negation symbol} 
$\neg$, \emph{and connectives} $\vee$ \emph{and} $\wedge$. \emph{Then, the}
selection of $R$ by $F$, denoted $\overline{\sigma}_F(R)$, \emph{is a 
generalized disjunctive paraconsistent relation on scheme} $\Sigma$,
\emph{given by} $\overline{\sigma}_F(R) = \gred(T)$, \emph{where}
$T^+ = \{w | w \in \gred(R)^+ \wedge ({\forall t} \in w)F(t)\}$ \emph{and}
$T^- = R^- \cup \sigma_{\neg F}(\tau(\Sigma))$, \emph{where} $\sigma_F$ \emph{is the usual selection of tuples}. 
\hfill{\space} $\Box$
\end{definition}

A disjunctive tuple set is either selected as a whole or not at all. All the
tuples within the tuple set must satisfy the selection criteria for the tuple
set to be selected.

\begin{definition}
\emph{Let} $R$ \emph{be a normalized generalized disjunctive paraconsistent 
relation on scheme} $\Sigma$ \emph{with} $\gred(R)^- = \{v_1, \ldots, v_n\}$., 
\emph{and} $\Delta \subseteq \Sigma$. 
\emph{Then, the} projection of $R$ onto $\Delta$, denoted $\overline{\pi}_\Delta(R)$, \emph{is a generalized disjunctive paraconsistent relation on scheme}
$\Delta$, \emph{given by} $\overline{\pi}_\Delta(R) = \gred(T)$, \emph{where}
$T$ \emph{is defined as follows}. \emph{Let} $E = \{\{t_1, \ldots, t_n\}|(\forall i)(1 \leq i \leq n \rightarrow t_i \in v_i)\}$. \emph{Let the elements of}
$E$ \emph{be} $E_1, \ldots, E_e$ \emph{and let} $A_i = \{t \in \pi(\Delta)|t^{\Sigma \cup \Delta} \subseteq (E_i)^{\Sigma \cup \Delta}\}$. Then, \\ 
$T^+ = \{\pi_\Delta(w) | w \in \gred(R)^+ \}$ \\
$T^- = \{w|(\exists t_1) \ldots (\exists t_e)(t_1 \in A_i \wedge \ldots \wedge t_e \in A_e \wedge w = \{t_1, \ldots, t_e\})\}$  
, \emph{where} $\pi_\Delta$ \emph{is the usual projection over} $\Delta$ \emph{of tuples}.
\hfill{\space} $\Box$
\end{definition}

The positive component of the projections consists of the projection of each of the tuple sets onto $\Delta$ and $\overline{\pi}_\Delta(R)^-$ consists of those
tuple sets in $\tau{(\Delta)}$, all of whose extensions are in $R^-$.

\begin{definition}
\emph{Let} $R$ \emph{and} $S$ \emph{be normalized generalized disjunctive 
paraconsistent relations on schemes} $\Sigma$ \emph{and} $\Delta$, \emph{respectively with} $\gred(R)^+ = \{v_1, \ldots, v_n\}$, \\ 
$\gred(R)^- = \{u_1, \ldots, u_k\}$ \emph{and} $\gred(S)^+ = \{w_1, \ldots, w_m\}$, \\ 
$\gred(S)^- = \{x_1, \ldots, x_j\}$. \emph{Then, the} natural join of $R$ \emph{and} $S$, \emph{denoted} $R \overline{\Join} S$, \emph{is a generalized disjunctive paraconsistent relation on
scheme} $\Sigma \cup \Delta$, \emph{given by} $R \overline{\Join} S = \gred(T)$,
\emph{where} $T$ \emph{is defined as follows. Let} 
$E = \{\{t_1, \ldots, t_n\} | (\forall i)(1 \leq i \leq n \rightarrow t_i \in v_i)\}$ \emph{and} $F = \{\{t_1, \ldots, t_m\} | (\forall i)(1 \leq i \leq m \rightarrow t_i \in w_i)\}$. \emph{Let the elements of} $E$ \emph{be} $E_1, \ldots, E_e$ \emph{and those of} $F$ \emph{be} $F_1, \ldots, F_f$ \emph{and let}
$A_{ij} = E_i \Join F_j$ \emph{for} $1 \leq i \leq e$ \emph{and} $1 \leq j \leq f$. \emph{Let} $A_1, \ldots, A_g$ \emph{be the distinct} $A_{ij}$s. 
\emph{Then,} \\
$T^+ = \{w | (\exists t_1) \cdots (\exists t_g)(t_1 \in A_1 \wedge \cdots \wedge t_g \in A_g \wedge w = \{t_1, \ldots, t_g\})\}$ \\
\emph{Let} $G = \{\{t_1, \ldots, t_k\}|(\forall i)(1 \leq i \leq k \rightarrow t_i \in u_i)\}$ \emph{and} $H = \{\{t_1, \ldots, t_j\}|(\forall i)(1 \leq i \leq j \rightarrow t_i \in x_i)\}$. \emph{Let the elements of} $G$ \emph{be}
$G_1, \ldots, G_g$ \emph{and those of} $H$ \emph{be} $H_1, \ldots, H_h$ \emph{and let} $B_{ij} = (G_i)^{\Sigma \cup \Delta} \cup (H_j)^{\Sigma \cup \Delta}$ 
\emph{for} $1 \leq i \leq g$ \emph{and} $1 \leq j \leq h$. \emph{Let} $B_1, \ldots, B_f$ \emph{be the distinct} $B_{ij}$s. \emph{Then}, \\
$T^- = \{w|(\exists t_1) \cdots (\exists t_f)(t_1 \in B_1 \wedge \cdots \wedge t_f \in B_f \wedge w = \{t_1, \ldots, t_f\})\}$.
\hfill{\space} $\Box$
\end{definition}

\begin{theorem}
Let $R$ and $S$ be two normalized generalized disjunctive paraconsistent relations on scheme $\Sigma_1$ and $\Sigma_2$. Also let $F$ be a selection formula on scheme $\Sigma_1$ and  $\Delta \subseteq \Sigma_1$. Then,

\vspace{-0.1in}

\begin{enumerate}
\item $\grep_{\Sigma_1}(\overline{\sigma}_F(R)) = {\cal S}(\widehat{\sigma}_F)(\grep_{\Sigma_1}(R))$.
\item $\grep_{\Sigma_1}(\overline{\pi}_\Delta(R)) = {\cal S}(\widehat{\pi}_\Delta)(\grep_{\Sigma_1}(R))$.
\item $\grep_{\Sigma_1 \cup \Sigma_2}(R \overline{\Join} S) = \grep_{\Sigma_1}(R) {\cal S}(\widehat \Join)\grep_{\Sigma_2}(S)$.
\end{enumerate}
\hfill{\space} $\Box$
\end{theorem}

\section{Conclusions and Future Work}

We have presented a framework for relational databases under which
positive disjunctive as well as explicit negative disjunctive facts can be
represented and manipulated. It is the generalization of disjunctive paraconsistent relation in \cite{SUN97}. There are at least two directions
for future work. One would be to make the model more expressive
by considering disjunctive positive and negative facts. 
Work is in progress in this
direction. 
The extended model will be more expressive
The algebraic operators
will have to be extended appropriately.
The other direction for future work would be to find
applications of the model presented in this paper.
There has been some interest in studying extended logic
programs
in which the head of clauses can have one or more literals
\cite{mnk93}. This
leads to two notions of negation: {\em implicit\/} negation 
(corresponding to negative literals in the body) and {\em explicit\/}
negation (corresponding to negative literals in the head).
The model presented in this paper could provide a 
framework under which the semantics of extended
logic programs could
be constructed in a bottom-up manner. 

\vspace{-.1in}


\vspace{-.1in}

\bibliography{/export/home/students/haibin/dasfaa05/ref}

\end{document}